%% file: main.tex
\documentclass[sigconf]{acmart}

\copyrightyear{2026}
\acmYear{2026}
\setcopyright{cc}
\setcctype{by}
\acmConference[MM '26] {Proceedings of the 34th ACM International Conference on Multimedia}{November 10--14, 2026}{Rio de Janeiro, Brazil.}
\acmBooktitle{Proceedings of the 34th ACM International Conference on Multimedia (MM '26), November 10--14, 2026, Rio de Janeiro, Brazil}
\acmISBN{979-8-4007-2213-4/2026/11}
\acmDOI{10.1145/767308.3835884}

\usepackage[svgnames]{xcolor}
\usepackage{amsmath} 
\usepackage{latexsym}
\usepackage{multirow} 
\usepackage{pifont}
\usepackage{float} 
\usepackage{placeins} 
\usepackage{fix-cm}
\usepackage{afterpage}
\usepackage{changepage}
\usepackage{subcaption} 

\usepackage{colortbl, xcolor}
\usepackage{makecell} 
\usepackage[table]{xcolor}  
\usepackage{booktabs}     
\usepackage{cuted}
\usepackage[hang,flushmargin]{footmisc}
\usepackage{siunitx}
\usepackage[most]{tcolorbox}
\usepackage{hyperref}

\begin{document}

\newtcolorbox{casestudy}[2][]{%
  colback=gray!1,
  colframe=gray!50,
  fonttitle=\bfseries,
  title={#2},
  breakable,
  enhanced,
  left=1em,
  right=1em,
  top=0.8em,
  bottom=0.8em,
}

\title[\textbf{\texttt{SpotSound}}]{
     \texttt{SpotSound}: Enhancing Large Audio-Language Models \\with Fine-Grained Temporal Grounding}

\author{Luoyi Sun}
\affiliation{
  \institution{Zhejiang University}
  \city{Hangzhou}
  \state{Zhejiang}
  \country{China}}

\author{Xiao Zhou}
\affiliation{
  \institution{SAI, Shanghai Jiao Tong University}
  \city{Shanghai}
  \country{China}}

\author{Zeqian Li}
\affiliation{
  \institution{SAI, Shanghai Jiao Tong University}
  \city{Shanghai}
  \country{China}}

\author{Ya Zhang}
\affiliation{
  \institution{SAI, Shanghai Jiao Tong University}
  \city{Shanghai}
  \country{China}}
  
\author{Yanfeng Wang}
\affiliation{
  \institution{SAI, Shanghai Jiao Tong University}
  \city{Shanghai}
  \country{China}}

\author{Weidi Xie}
\affiliation{
  \institution{SAI, Shanghai Jiao Tong University}
  \city{Shanghai}
  \country{China}}

\definecolor{b1}{HTML}{1D4ED8}
\definecolor{r1}{HTML}{a52a2a}
\renewcommand{\shortauthors}{Luoyi Sun et al.}

\input{sections/0_abstract}





\maketitle
\input{sections/1_intro}

\input{sections/3_method}
\input{sections/4_dataset}
\input{sections/5_experiments}

\input{sections/2_related_works}
\input{sections/6_conclusion}

\begin{acks}
This work is supported by the Scientific Research Innovation Capability Support Project for Young Faculty (ZY-GXQNJSKYCXNLZCXM-I22) and the Shanghai Special Fund for Promoting High-Quality Industrial Development: Pilot Industry Innovation and Development Program (Artificial Intelligence Track, ID: 2025-GZL-RGZN-02020).
\end{acks}

\bibliographystyle{ACM-Reference-Format}
\bibliography{sample-base-new}

\appendix
\vspace{3pt} \begin{center} \noindent   \textbf{\Large Appendix}  \end{center} \vspace{2pt}
\input{sections/9_appendix}
\end{document}

%% file: sections/0_abstract.tex
\begin{abstract}
Large Audio-Language Models (ALMs) have recently demonstrated remarkable capabilities in holistic audio understanding, yet they remain unreliable for temporal grounding, {\em i.e.}, the task of pinpointing exactly when an event occurs within long-form audio. This limitation stems from two factors: training data dominated by clip-level supervision lacking precise timestamps, and benchmarks that fail to simulate real-world scenarios where short events are obscured by dense background sounds. In this paper, we introduce \textbf{\texttt{SpotSound}}, 
an audio language model designed for grounding audio events. \textbf{\texttt{SpotSound}} incorporates a novel training objective, specifically designed to suppress hallucinated timestamps for events absent from the input. Additionally, we present \textbf{\texttt{SpotSound-Bench}}, a challenging temporal grounding benchmark where target events occupy less than ~10\% of each clip, creating a rigorous `needle-in-a-haystack' evaluation. Experiments demonstrate that \textbf{\texttt{SpotSound}} achieves state-of-the-art results on temporal grounding benchmarks while maintaining robust performance across general downstream audio-language tasks. Code, models and benchmark are released on https://loiesun.github.io/spotsound/
\end{abstract}

%% file: sections/1_intro.tex
\section{Introduction}
\label{sec:intro}

Large Audio-Language Models (ALMs)~\cite{chu2024qwen2, goel2025audio,ding2025kimi} have recently demonstrated remarkable proficiency in holistic tasks, such as generating captions or summarizing entire audio clips. However, they exhibit deficiencies in temporal grounding: the ability to precisely localize specific events within a continuous audio stream. This limitation restricts their deployment in practical applications like security surveillance and media forensics, where precise timing is as crucial as the event classification itself.

\begin{figure}[t]
\centering
\includegraphics[width=\linewidth]{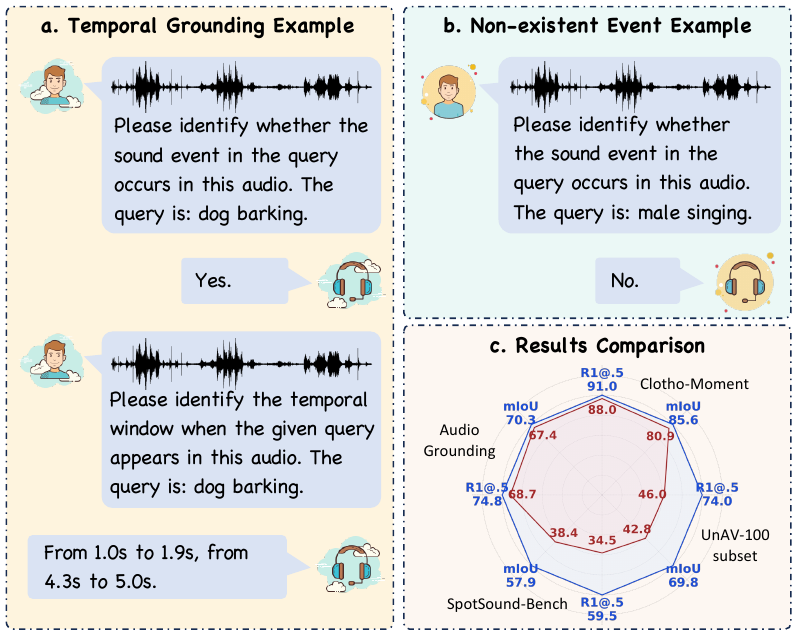}
\vspace{-0.6cm}
\caption{Qualitative examples and performance comparison.
(a) \textbf{\texttt{SpotSound}} accurately grounds timestamps relevant to the query.
(b) \textbf{\texttt{SpotSound}} identifies sound events described in the query that are absent from the audio.
(c) Quantitative comparison; \textcolor{b1}{blue} denotes \textbf{\texttt{SpotSound}}, and \textcolor{r1}{red} denotes previous top-tier models. We evaluate models on four benchmarks and report the mIoU and R1@.5 for each.}
\vspace{-0.3cm}
\label{fig:arch}
\end{figure}

Generally speaking, two primary factors impede the temporal grounding capabilities of current models: (i) the majority of large ALMs~\cite{ding2025kimi, bai2023qwen,kong2024audio} are trained on data with coarse, `clip-level' annotations, the models therefore learn to associate acoustic events with the full audio duration rather than precise temporal boundaries; (ii) the community lacks challenging benchmarks to rigorously measure progress for grounding. Existing datasets~\cite{xu2021text, munakata2025language, geng2023dense} predominantly feature distinct, long-duration sound events separated by silence, targets that are relatively trivial to locate. In contrast, real-world audio is characterized by complex acoustic scenes where short, fleeting events are embedded within continuous background noise, making detection significantly more challenging.

\begin{figure*}[htbp]
\centering
\includegraphics[width=\linewidth]{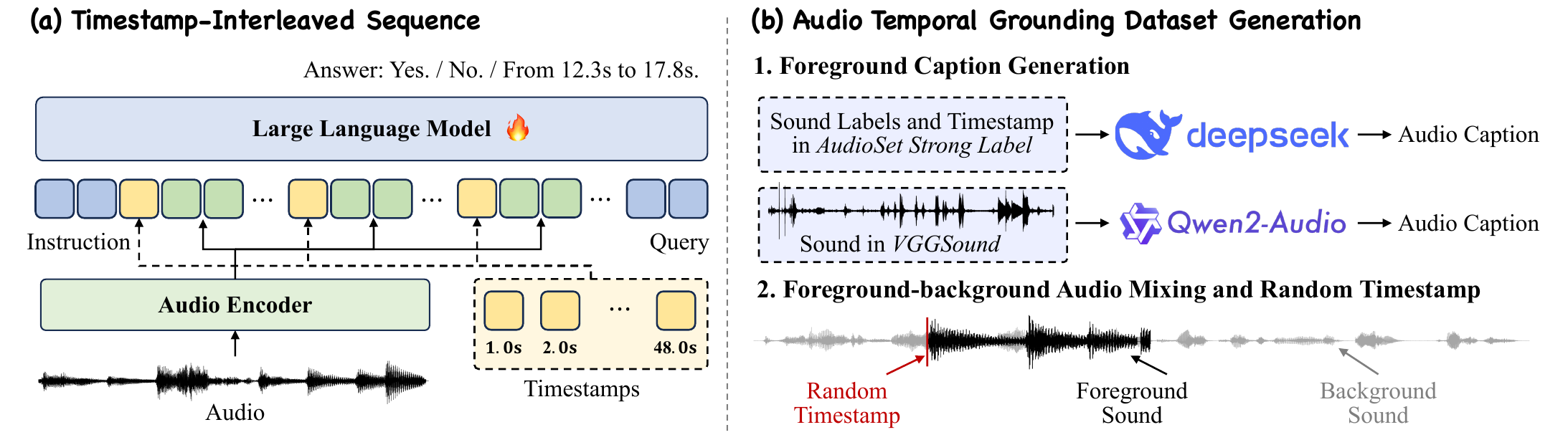}
\vspace{-0.9cm}
\caption{Model architecture and dataset generation pipeline. 
(a) In \textbf{\texttt{SpotSound}}, we construct an interleaved sequence of timestamps and audio tokens and concatenate with a language query, and feed it to the LLM to predict the target temporal interval. 
(b) We employ LLMs to generate captions for foreground audio, then randomly mix foreground and background sounds to synthesise the final audio, preserving the insertion timestamp as ground-truth. }
\vspace{-0.2cm}
\label{fig:arch}
\end{figure*}

This paper presents \textbf{\texttt{SpotSound}}, a framework designed to bridge this gap by equipping large ALMs with precise temporal reasoning capabilities. Central to our approach is a temporal encoding mechanism that interleaves timestamp tokens with audio embeddings, enabling accurate event boundary localization for open-vocabulary natural language queries. To ensure response reliability, we introduce a specialized training objective that explicitly targets hallucination mitigation. Specifically, we restructure each training instance into a discriminative quadruplet format, comprising the audio input, a positive query~(describing a present sound event), the corresponding ground-truth event timestamps, and a negative query~(describing an absent event). 
This formulation enforces the model to verify acoustic evidence, learning to distinguish genuinely occurring signals from non-existent ones. To support end-to-end training, we curate a temporally-aware audio-language dataset comprising 10k instruction-tuning samples, and collect an additional 67.6k samples from existing audio-language datasets, resulting in 77.6k samples in total. 

In addition, we address the evaluation gap by introducing a benchmark, 
termed \textbf{\texttt{SpotSound-Bench}}, which is specifically designed for short-window temporal audio grounding. Unlike prior benchmarks~\cite{xu2021text, munakata2025language, geng2023dense} that focus on distinct, isolated sound events, \textbf{\texttt{SpotSound-Bench}} embeds short target windows within long-form audio, creating a needle-in-a-haystack scenario. The target events occupy less than 10\% of the total duration, forcing the model to pick out fleeting acoustic cues against rich, competing background activity. This makes the benchmark a demanding testbed for evaluating temporal grounding under realistic, dense audio conditions.

In summary, we make the following contributions:
(i) we endow large ALMs with robust temporal grounding via a training objective that directly suppresses hallucinations on non-existent events;
(ii) we incorporate temporal information by interleaving timestamp tokens with audio tokens, giving the model the fine-grained resolution needed to capture the precise timing and duration of acoustic events;
(iii) we construct a challenging temporal audio grounding benchmark, \textbf{\texttt{SpotSound-Bench}}, filling a gap in evaluation resources for long-duration, realistic temporal reasoning;
(iv) extensive experiments confirm state-of-the-art performance across multiple temporal grounding benchmarks, with the model retaining strong accuracy on the standard sound event detection (SED) task, indicating broad generalization.

%% file: sections/3_method.tex
\section{Methods}
\label{sec:methods}

In this section, we present \textbf{\texttt{SpotSound}}, a framework that endows large Audio-Language Models~(ALMs) with the ability of precise temporal grounding across varying audio durations. 
We start with the problem formulation in Section~\ref{sec:problem}, 
and detail our architectures and the construction of timestamp-interleaved sequences in Section~\ref{sec:atgm}.
Lastly, we outline the training paradigm in Section~\ref{sec:training}.

\subsection{Problem Formulation}
\label{sec:problem}
Given an audio stream $\mathcal{A} = \{a_1, \ldots, a_{T}\} \in \mathbb{R}^{1 \times T}$ and a free-form textual query  \(\mathcal{Q}\), either a concise label or a descriptive caption, we treat the temporal grounding as a two-stage problem, where the model first answers a binary existence instruction $\mathcal{I}_E$ by predicting:
\[
\mathcal{B} = \Phi_{\text{SpotSound}}(\mathcal{A}, \mathcal{I}_E)
\tag{1}
\]
where $\mathcal{B} = \{\mathbf{yes}$, $\mathbf{no}$\}. 
If $\mathcal{B} = \mathbf{yes}$, the model proceeds to the second stage and answers a grounding instruction, noted as $\mathcal{I}_G$, by localizing all time intervals that semantically match the query:
\[
\mathcal{W} = \Phi_{\text{SpotSound}}(\mathcal{A}, \mathcal{I}_G)
\tag{2}
\]
where $\mathcal{W} = \{(s_1, e_1), \ldots, (s_K, e_K)\}$ correspond semantically to the query, where each pair $(s_k, e_k)$ represents a respective start and end timestamp.

\subsection{Audio Temporal Grounding Model}
\label{sec:atgm}

In this section, we provide details of the proposed \textbf{\texttt{SpotSound}} model for audio temporal grounding that localizes time spans in an audio recording corresponding to a natural-language query. 

\vspace{3pt} \noindent \textbf{Large Audio Language Backbone.}
Contemporary large ALMs typically couple an audio encoder with a large language model (LLM). In this study, we adopt the two representative models, Qwen2-Audio~\cite{chu2024qwen2} and Audio Flamingo 3~\cite{goel2025audio}, as our backbone models.
 
Both of these backbones initialize Whisper-large-v3~\cite{radford2023robust} as audio encoder.
In this process, raw audio $\mathcal{A}$ is initially resampled to 16kHz and subsequently converted into 128-channel mel-spectrograms  $\mathcal{M} = \{m_1, \ldots, m_{T}\}\in \mathbb{R}^{T_{\text{mel}} \times F}$ using a 25ms window and a 10ms hop length, where $T_{\text{mel}}$ and  $F$ represent the time and frequency channels.
The spectrograms are subsequently encoded into audio tokens 
via the audio encoder, \(\mathbf{A}_i = \phi_{\text{audio}}(m_i)\). The encoder incorporates a pooling layer with a stride of two, compressing the temporal length of the audio representation. 
Each output timestep of the encoder corresponds to approximately 40ms of the original audio signal.

Qwen2-Audio~\cite{chu2024qwen2} establishes a strong baseline by leveraging Qwen2-7B~\cite{yang2024qwen2technicalreport} as its language model, while Audio Flamingo 3~\cite{goel2025audio} adopts the subsequent Qwen2.5-7B~\cite{qwen2025qwen25technicalreport} iteration.

\vspace{3pt} \noindent \textbf{Timestamp-Interleaved Sequence Construction.}
To establish a precise alignment between audio features and their temporal positions, we explicitly encode time by inserting textual timestamp tokens before the corresponding audio tokens at a fixed granularity. As illustrated in Figure~\ref{fig:arch}(a), for each time index $t_i$, we construct a textual timestamp token $\tau_i=\texttt{``timestamp: }t_i\texttt{ seconds''}$ and place it immediately before the corresponding audio frame features, and we set the granularity of timestamps to 1 second. 
This yields the interleaved sequence:
\[
S = [\mathbf{T}_1; \mathbf{A}_1; \mathbf{T}_2; \mathbf{A}_2; \ldots; \mathbf{T}_{n}; \mathbf{A}_{n}; \mathbf{I}; \mathbf{Q}],
\vspace{-3pt}
\tag{3}
\label{eq:seq_format} 
\]
\(\mathbf{T}_i = \phi_{\text{tokenizer}}(\tau_i)\), 
\(\mathbf{Q} = \phi_{\text{tokenizer}}(\mathcal{Q} )\), \(\mathbf{I} = \phi_{\text{tokenizer}}(\mathcal{I} )\), 
where \(\phi_{\text{tokenizer}}\) denotes the language tokenization in \( \Phi_{\text{SpotSound}} \), \(\mathcal{I}\) can be \(\mathcal{I}_E \) or  \(\mathcal{I}_G \),
and $n$ represents the duration of the audio. 
This interleaved sequence is then fed into the large language model (LLM), which generates temporal boundaries for a given query \(\mathcal{Q}\), with the output formatted as \(\mathcal{\hat{B}} =\)``\texttt{Yes.}'' or ``\texttt{No.}'', or alternatively
\(\mathcal{\hat{W}} =\)``\texttt{From \(s_k\) seconds to \(e_k\) seconds}''. 
In essence, we harness the retrieval capabilities of ALMs to read out the inserted timestamp tokens, rather than decoding dense positional encodings.

\subsection{Training Strategy}
\label{sec:training}
We train the model for temporal grounding with an auto-regressive objective. For each training instance \( (\mathcal{A}, \mathcal{I}, \mathcal{Q}, \mathcal{Y}) \), namely, the audio input, the instruction, the natural-language query, and the target output, the prompt sequence \( S \)~(as shown in Eq.~\eqref{eq:seq_format}) interleaves textual tokens (including the query, the instruction, and timestamp tokens) with audio tokens. The model is trained by minimizing the negative log-likelihood over the target tokens exclusively:
\[
\mathcal{L} = -\sum_{i=1}^{N_y} \log P(y_i \mid S, y_{<i}; \theta),
\tag{4}
\]
where \( N_y \) is the target length. $y_{<i}$ denotes all target tokens preceding the $i$-th token, \( \mathcal{Y} = \mathcal{B}\), if \( \mathcal{I} = \mathcal{I}_E\), and \( \mathcal{Y} = \mathcal{W}\), if \( \mathcal{I} = \mathcal{I}_G\).

%% file: sections/4_dataset.tex
\section{Training Dataset and Benchmark}
\label{sec:databench}
Here, we detail the dataset used for training, and our proposed short-window sound event benchmark, \textbf{\texttt{SpotSound-Bench}}. 

\subsection{Training Dataset}
\label{sec:data}

We begin with an analysis of existing datasets, describe our synthetic data pipeline and construction of negative samples. To support joint training, we assemble a diverse corpus by combining publicly available datasets with a newly generated set rich in dense linguistic annotations. In total, the corpus contains 77.6k samples spanning a wide range of audio durations and query formats.

\vspace{3pt} \noindent \textbf{Existing Datasets.} 
Fine-grained audio understanding remains constrained by the scarcity of high-quality, temporally aligned annotations. 
To address this, we construct a unified training set from several existing datasets, as summarized in Table~\ref{tab:dataset}. Specifically, we draw on temporal grounding datasets, \textit{e.g.}, AudioGrounding~\cite{xu2021text} and Clotho-Moment~\cite{munakata2025language}, alongside densely time-stamped classification corpora, namely UnAV-100~\cite{geng2023dense} and AudioSet Strong Label, denoted as ASSL~\cite{hershey2021benefit}. For ASSL, we randomly select 5,000 clips from its training split to maintain a balanced composition.

These sources vary considerably in scope and character: (i) annotations are either human-curated or automatically generated; (ii) recordings range from short clips to long-form audio; and (iii) queries are either caption-driven (free-form natural language) or label-centric (fixed-vocabulary event identifiers).
To unify them under a common training objective, we convert each sample into a standardized format of a textual query paired with a (\textit{start}, \textit{end}) timestamp, where timestamps are uniformly represented with two decimal places to ensure fine-grained temporal precision. For label-centric datasets such as UnAV-100 and ASSL, we directly use the original event labels as textual queries. For caption-driven datasets, including AudioGrounding and Clotho-Moment, we retain the original captions or moment queries. All data are sourced from the official training splits, preserving their original partitions.

\begin{table}[t]
\setlength{\tabcolsep}{4.5pt} 
\centering
\caption{\textbf{Statistics of the training set. } `A\_num' and `Q\_num' denote the total number of audio clips and queries, respectively. `Anno', `DUR', and `Q\_type' represent the annotation method, average duration, and query type. Under annotation, `M.' indicates manual annotation, while `A.' indicates automatic annotation. Under query type, `Cap.' indicates caption.}
\vspace{-0.3cm}
{
\begin{tabular}{lccccc}
\toprule
Datasets & A\_num & Q\_num & Anno. & DUR. & Q\_type\\
\midrule
AudioGrounding & 3,770 & 8,935 & M. & 10s  & Cap.\\
Clotho-Moment  & 32,694 & 32,694 & A. & 60s & Cap. \\
UnAV-100 & 5,686 & 9,115 & M. & 43s  & Label \\
ASSL & 5,000 & 16,896 & M. & 10s  & Label\\
\midrule
Ours & 10,000 & 10,000 & A. & 50s  & Cap. \\
\bottomrule
\end{tabular}
}
\vspace{-0.2cm}
\label{tab:dataset}
\end{table}

\vspace{3pt} \noindent \textbf{Long-form Synthetic Dataset.}
While the datasets above are valuable, their queries tend to be sparse, with captions that often reduce to little more than simple event labels. To enrich supervision with denser linguistic cues, we construct a novel long-form dataset featuring detailed, temporally grounded audio captions.

We randomly sample 5,000 clips from the strongly labelled subset of AudioSet~\cite{gemmeke2017audio, hershey2021benefit} and 5,000 clips from VGGSound~\cite{chen2020vggsound} as foreground events. As illustrated in Figure~\ref{fig:arch}(b), we generate fine-grained, temporally aware captions from ASSL's multi-segment, multi-label annotations using DeepSeek-v3~\cite{liu2024deepseek}. For VGGSound clips, we prompt Qwen2-Audio~\cite{chu2024qwen2} to produce vivid, descriptive captions tailored to each clip's content. 
To improve foreground salience and remove dead time, we apply dataset-specific trimming strategies to the foreground clips. For AudioSet samples, we rely on the strong annotations: we merge the time intervals of all constituent events into a continuous segment spanning from the earliest start time to the latest end time, discarding any audio portions outside this merged range. For VGGSound samples, we trim leading and trailing silence based on signal energy, specifically removing segments more than 20 dB below the mean signal power, and retain the remaining continuous audio segment.

We then sample a clip (40-60s) from Walking Tours~\cite{venkataramanan2023imagenet} as background ambience. 
Each trimmed foreground segment is then randomly placed into the background, with its start position sampled uniformly across the background duration. The final timestamps of the query are thus determined by this placement: the start time is the sampled insertion point, and the end time is the insertion point plus the duration of the trimmed foreground.
Mixing levels are randomized: foreground gain is jittered by $\pm$5 dB relative to its nominal level, and background gain is fixed at $-10 \pm 5$ dB relative to the foreground, producing varied signal-to-noise conditions.

For the ASSL and VGGSound datasets, we randomly sampled 100 model-generated audio captions each and manually verified their correspondence with the original audio, yielding an accuracy above 95\% for both. Additional dataset statistics on sound events are provided in the Appendix~\ref{sec:statisd} 

\vspace{3pt} \noindent \textbf{Negative Samples.}
To improve robustness against hallucinations, {\em i.e.}, predicting whether an event actually presents, 
we pair each training sample with a negative counterpart. For a given audio clip, the associated caption or label serves as the positive query, while a negative query is drawn from events absent from that clip.

Concretely, we pool all queries across the training corpora into a global query set. 
For each clip, a negative query is sampled subject to two constraints: (i) it does not appear in the clip's annotations, and (ii) it shares no lexical overlap with the positive query, reducing the chance of spurious matches. Each audio instance is thus paired with two question types: a presence question, asking whether the described event occurs in the audio, answered \textit{yes} or \textit{no}, and a localization question, asking for the temporal interval of the event, answered with the relevant time window.

\vspace{-0.35cm}
\subsection{Benchmark}
We evaluate our model across multiple benchmarks under varying configurations, 
{\em e.g.}, varying audio durations and window lengths.

\vspace{3pt} \noindent \textbf{Existing Benchmarks.}
As summarized in Table~\ref{tab:bench}, we evaluate on AudioGrounding~\cite{xu2021text}, Clotho-Moment~\cite{munakata2025language}, and the UnAV-100 subset~\cite{munakata2025language}, a manually annotated subset of the UnAV-100 test set. 
These benchmarks span two temporal regimes: AudioGrounding is a short-form benchmark that probes localization of transient events in short clips, whereas Clotho-Moment and UnAV-100 subset are long-form benchmarks that assess localization of sustained time spans in extended, untrimmed recordings.

\vspace{3pt} \noindent 
\textbf{\texttt{SpotSound-Bench}}.
A persistent limitation of existing benchmarks is the high ratio of target-window duration to the full audio clip. The average coverage is 26\% on AudioGrounding, 33\% on Clotho-Moment, and 28\% on UnAV-100 subset, which effectively narrows the search space and simplifies the task. Here, we present a benchmark that features short acoustic events embedded within long, unstructured recordings. Specifically, using YouTube as the data source, we retrieve and annotate in-the-wild audio guided by the 100-category ontology of UnAV-100~\cite{geng2023dense}, and we collect long-form audio, focusing on short-window grounding. 

As summarized in Table~\ref{tab:bench}, our proposed benchmark contains 400 audio-query-timestamp triplets. The average clip is 54.2s, with target events averaging 3.9s, yielding a temporal density of 7.2\%. This design creates a large search space dominated by background content and demands high temporal precision from models. We release \textbf{\texttt{SpotSound-Bench}},   including audio streams and timestamp annotations to facilitate reproducible evaluation.  
\begin{table}[t]
\setlength{\tabcolsep}{4pt} 
\centering
\caption{
\textbf{Benchmark statistics. } `A\_num' and `Q\_num' denote the total number of audio clips and queries, respectively. `DUR', `W\_len' and `Q\_type' represent the average duration, average window length and query type. Under query type, `Cap.' indicates caption.}
\vspace{-0.15cm}
{
\begin{tabular}{lccccc}
\toprule
Models & A\_num & Q\_num &  DUR. & W\_len & Q\_type\\
\midrule
AudioGrounding  & 70 & 100 & 10s & 2.6s & Cap.  \\
Clotho-Moment  & 6,649 & 6,649  & 60s & 19.6s & Cap. \\
UnAV-100 subset & 492 & 997 & 42.4s  & 14.6s & Cap. \\
\midrule
\textbf{\texttt{SpotSound-Bench}} & 400 & 400 & 54.2s & 3.9s  & Label \\
\bottomrule
\end{tabular}
}
\vspace{-0.3cm}
\label{tab:bench}
\end{table}

\vspace{3pt} \noindent \textbf{Negative Samples.} 
Following the same setup as the training data, for each audio clip in the benchmarks, we construct a positive query and a negative query, to evaluate whether the model could correctly determine the presence of the sound event. Additional benchmark statistics on sound events are provided in the Appendix~\ref{sec:statisb}

%% file: sections/5_experiments.tex
\section{Experiments}
\label{sec:experiments}

In Section~\ref{sec:atg}, we evaluate \textbf{\texttt{SpotSound}} on  Audio Temporal Grounding (ATG), and compare its performance with existing methods across multiple benchmarks. In Section~\ref{sec:hallu}, we further investigate the ability to determine whether a target event is present. In Section~\ref{sec:twostage}, we compare the performance with existing models by combining the two stages. Furthermore, in Section~\ref{sec:sed}, we evaluate on Sound Event Detection (SED) benchmarks to validate the generalization capability. Finally, Section~\ref{sec:abla} provides ablation studies on key components and hyperparameters.

\begin{table*}[]
\setlength{\tabcolsep}{5.8pt} 
\centering
\caption{Audio temporal grounding results. mIoU represents the mean IoU. R1@.3 and R1@.5 denote Recall@1 across IoU thresholds of 0.3 and 0.5, respectively. \texttt{\textbf{SpotSound-Q}} and \texttt{\textbf{SpotSound-A}} denotes the integration of the SpotSound method with Qwen2 Audio and Audio Flamingo 3, respectively. Best results are in bold, and second-best are \underline{underlined}.}
\vspace{-0.15cm}
{
\begin{tabular}{lccc|ccc|ccc|ccc}
\toprule
\multirow{2}{*}{Models} & \multicolumn{3}{c|}{Clotho-Moment} & \multicolumn{3}{c|}{UnAV-100 subset}  & \multicolumn{3}{c|}{\texttt{\textbf{SpotSound-Bench}}} & \multicolumn{3}{c}{AudioGrounding}\\
\cline{2-13}
&   R1@.3 & R1@.5 & mIoU & R1@.3 & R1@.5  & mIoU & R1@.3 & R1@.5  & mIoU & R1@.3 & R1@.5  & mIoU \\
\midrule
\rowcolor{gray!10}\multicolumn{13}{c}{\textit{Non-LLM Models}} \\
WTATG~\cite{xu2024towards}    &12.1 &6.3 & 9.1   & 53.0 & 37.0 & 38.4 & 53.0& 34.5 & 38.4 & 72.5 & 53.7  & 51.4\\
AM-DETR~\cite{munakata2025language}    & 89.8 & 88.0 & 80.9 & 59.0 & 46.0 & 42.8 & 25.5 & 15.0 & 19.5 &52.5 & 15.6  & 30.2\\
\midrule
\rowcolor{gray!10}\multicolumn{13}{c}{\textit{Proprietary Models}} \\
Gemini-2.5-Flash~\cite{comanici2025gemini}& 45.7& 33.7 & 36.9 & 48.0 & 37.0 & 35.6 & 34.8 &  27.8 & 25.7 & 51.8 & 36.2& 37.1\\
Gemini-2.5-Pro~\cite{comanici2025gemini} & 40.4 & 32.5 & 32.5 & 39.0 & 35.0 & 34.6 & 20.0 & 17.3 & 19.7 &  45.3 & 31.7 & 33.5\\
\midrule
\rowcolor{gray!10}\multicolumn{13}{c}{\textit{Open-Source Models}} \\
Kimi-Audio~\cite{ding2025kimi}   & 0.7 & 0.1 & 0.9 & 8.0 & 2.0 & 5.3 & 3.0 & 1.3 & 3.2 & 4.6 & 2.8 & 4.9 \\
TimeAudio~\cite{wang2025timeaudio}    & 39.5 & 24.9 & 28.6 & 21.0 & 7.0 & 16.0 & 7.8 & 1.8& 10.2 & 83.3 & \underline{68.7}  & 67.4\\
 Qwen2-Audio~\cite{chu2024qwen2}    & 6.1 & 1.3 & 5.7 & 14.0 & 4.0 & 9.7 & 5.5 & 1.8 & 4.4 & 50.3 & 29.3  & 37.0\\

\rowcolor{blue!5} \textbf{\texttt{SpotSound-Q}} & \textbf{93.6} & \textbf{91.2} & \underline{85.4} & \textbf{88.0} & \textbf{74.0} & \textbf{72.4} & \underline{65.0} & \underline{51.3} & \underline{49.9} & \underline{87.2} & {66.1 } & \underline {67.8}  \\

 Audio Flamingo 3~\cite{goel2025audio}    & 32.9 & 21.8 & 22.6 & 35.0 & 24.0 & 25.0 & 8.8 & 2.8 & 7.6 & 66.9 & 42.0  & 47.5\\
\rowcolor{blue!5} \textbf{\texttt{SpotSound-A}} & \underline{93.4} & \underline{91.0} & \textbf{85.6} & \underline{86.0} & \textbf{74.0} & \underline{69.8} & \textbf{76.5} & \textbf{59.5} & \textbf{57.9} & \textbf{90.1} & \textbf{74.8 } & \textbf{70.3}  \\

\bottomrule
\end{tabular}
}
\label{tab:atg}
\end{table*}

\subsection{Audio Temporal Grounding}
\label{sec:atg}
We evaluate \textbf{\texttt{SpotSound}} comparing against two task-specific methods, WTATG~\cite{xu2024towards} and AM-DETR~\cite{munakata2025language}, as well as recent large audio-language models (ALMs), including Kimi-Audio~\cite{ding2025kimi}, Audio Flamingo 3~\cite{goel2025audio}, TimeAudio~\cite{wang2025timeaudio}, and Qwen2-Audio~\cite{chu2024qwen2}.

\vspace{3pt} \noindent \textbf{Implementation Details.} 
Our framework adopts Qwen2-Audio~\cite{chu2024qwen2} and Audio Flamingo 3~\cite{goel2025audio} as backbone models. 
Qwen2-Audio processes at most 30 seconds of audio per forward pass; for longer recordings, we partition the input into contiguous 30-second segments, encode each independently, and concatenate the resulting features in temporal order into a unified sequence.
All experiments use the AdamW optimizer~\cite{loshchilov2017decoupled} with a learning rate of 1e-4, trained for one epoch with a linear warmup over the first 1,000 steps. The audio encoder is kept frozen throughout training, while the LLM is fine-tuned via LoRA~\cite{hu2022lora} with rank 8 and alpha 16.

\vspace{3pt} \noindent \textbf{Metrics.}
We first apply regular expressions to extract timestamp values from free-text responses \(\mathcal{\hat{W}} \), then arrange these into paired time windows \(\mathcal{\hat{W}}_\text{p}\). We adopt Recall@1 (R1) at different Intersection over Union (IoU) thresholds $\theta$, together with mean IoU (mIoU), as our evaluation metrics across all benchmarks. Specifically‌,
==\[R1@\theta = \frac{1}{N} \sum_{i=1}^{N}  \mathbf{1}_{\left( \frac{|\mathcal{W}_i \cap \mathcal{\hat{W}}_{\text{p}i}|}{|\mathcal{W}_i \cup \mathcal{\hat{W}}_{\text{p}i}|} \ge \theta \right)}, \quad   \text{mIoU} = \frac{1}{N} \sum_{i=1}^{N} \text{IoU}(\mathcal{W}_i, \mathcal{\hat{W}}_{\text{p}i}),\]
where $N$ indicates the number of data pairs in benchmark, and \(\theta \in \{0.3, 0.5\}\).

\vspace{3pt} \noindent \textbf{Results.}
As shown in Table~\ref{tab:atg}, 
our comparisons highlight key limitations of prior approaches and the efficacy of our method. Built upon Qwen2-Audio~\cite{chu2024qwen2} and Audio Flamingo 3~\cite{goel2025audio}, we developed \texttt{\textbf{SpotSound-Q}} and \texttt{\textbf{SpotSound-A}}.


First, task-specific models generalize poorly across distributions. WTATG, for instance, achieves a mIoU of 51.4 on its training benchmark AudioGrounding, yet collapses to 9.1 on Clotho-Moment. However, it is worth noting that these specialized models retain a degree of temporal stability on complex audio; their performance on \texttt{\textbf{SpotSound-Bench}} exceeds that of existing ALMs.

Second, existing large audio-language models, {\em e.g.}, Kimi Audio, Qwen2-Audio, demonstrate strong semantic understanding but struggle with precise temporal localization, yielding weak performance across benchmarks. Audio Flamingo 3 shows partial temporal competence but still struggles on challenging cases (mIoU 7.6 on \textbf{\texttt{SpotSound-Bench}}). Models like TimeAudio, even trained with temporal annotations ({\em e.g.}, AudioGrounding), perform well on short-clip benchmarks (67.4 mIoU on AudioGrounding), but generalize poorly to long or complex recordings. In addition, we evaluate the proprietary models, Gemini-2.5-flash and Gemini-2.5-pro, both of which demonstrate only basic temporal localization ability, with mIoU scores consistently below 40 across all benchmarks.

Third, our approach is designed to be compatible with any large audio-language model based on LLMs, and we apply our method to different backbone models, including Qwen2-Audio and Audio Flamingo 3, as \texttt{\textbf{SpotSound-Q} and \texttt{\textbf{SpotSound-A}}}. Our models achieve state-of-the-art or competitive results across all benchmarks, where \texttt{\textbf{SpotSound-A}} surpasses previous methods in mIoU: Clotho-Moment~(+4.7\%), UnAV-100 subset~(+27.0\%), AudioGrounding~(+2.9\%), and \textbf{\texttt{SpotSound-Bench}}~(+19.5\%). Note that, the gains on our proposed benchmark are notable and provide evidence of superior temporal precision, as it is carefully annotated. These results demonstrate the generalization ability of our method, and indicate that its effectiveness can benefit from advances in large ALMs.
With the emergence of more powerful large ALMs,
we expect further enhancements in temporal grounding performance.

Overall, our model generalizes across diverse audio domains and excels at fine-grained grounding of short, text-described acoustic events within complex auditory scenes. Furthermore, we provide failure cases and analysis in the Appendix D.

\begin{figure*}[t]
\centering
\includegraphics[width=\linewidth]{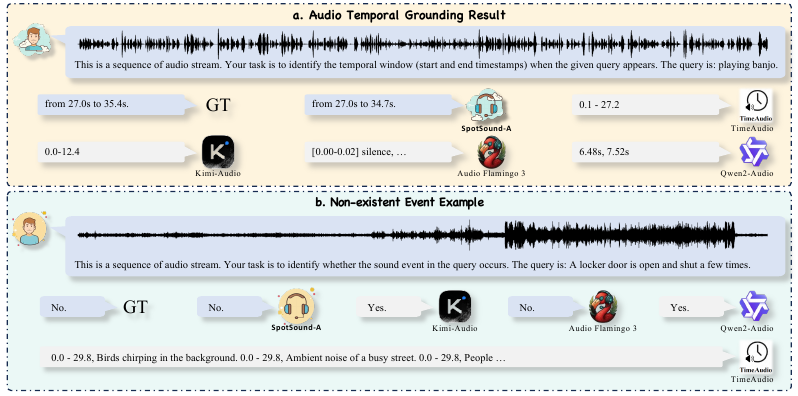}
\vspace{-0.7cm}
\caption{Qualitative comparison with other large audio-language models.  (a) Example cases from the \textbf{\texttt{SpotSound-Bench}}, illustrating the temporal windows predicted by each model in response to the textual query.  (b) Examples drawn from the Clotho-Moment, demonstrating cases where the models identify the non-existent sound events. }
\vspace{-0.1cm}
\label{fig:result}
\end{figure*}

\vspace{3pt} \noindent \textbf{Qualitative Results.}
As illustrated in Figure~\ref{fig:result}a, a representative sample from \textbf{\texttt{SpotSound-Bench}} highlights the superior temporal precision of our approach. \textbf{\texttt{SpotSound-A}} achieves the highest grounding accuracy with a predicted IoU of 91.7\%.
In contrast, existing large ALMs exhibit distinct failure modes when confronted with fine-grained temporal tasks. Qwen2-Audio generates syntactically complete time windows that suffer from severe semantic misalignment, resulting in zero overlap with the ground truth. Conversely, Kimi-Audio and TimeAudio demonstrate a ``collapse'' behavior under high uncertainty, defaulting to trivial intervals starting at $0$. Finally, Audio Flamingo 3 struggles with complex acoustic scenes, exhibiting autoregressive degradation where it iteratively generates erroneous, frame-by-frame hallucinations.
These comparisons underscore our model's robustness in bridging the gap between audio semantics and precise temporal boundaries.

\subsection{Hallucination for Negative Samples}
\label{sec:hallu}
A recurring failure mode in previous temporal grounding models built upon Multimodal Large Language Models (MLLMs)~\cite{wang2025timeaudio} is the tendency to predict temporal windows regardless of whether the queried event is actually present in the audio. To investigate this, we design a controlled experiment in which models are given an unrelated query and asked to determine whether the described sound event occurs in the clip.

\vspace{3pt} \noindent \textbf{Benchmarks.}
We construct negative samples from four benchmarks, Clotho-Moment, AudioGrounding, UnAV-100 subset and \texttt{\textbf{SpotSound-Bench}}. For each entry, we retain the original audio and replace its query with a sound description absent from the clip, forming a matched negative pair. The original query-audio pair serves as the corresponding positive sample.

\vspace{3pt} \noindent \textbf{Metrics.}
We evaluate binary event presence prediction to quantify hallucination behavior: a model is counted as correct on a negative sample if it predicts the queried event as absent. Accuracy is reported over the combined positive and negative sets.

\vspace{3pt} \noindent \textbf{Results.}
As demonstrated in Table~\ref{tab:hall}, we can make three observations:
(i) existing large audio-language models show nontrivial ability to judge event presence. Audio Flamingo 3 performs best, reaching 89.1\% accuracy on positives and 76.0\% on negatives for AudioGrounding;
(ii) TimeAudio, a representative audio-language model for audio temporal understanding, 
exhibits pronounced hallucination, which tends to output time spans even when the queried event is absent;
(iii) our models reliably determine event presence, demonstrating strong robustness. Relative to Audio Flamingo 3, it improves average existence accuracy by +18.8\% on Clotho-Moment and +8.1\% on AudioGrounding. More results of UnAV-100 subset and \texttt{\textbf{SpotSound-Bench}} are provided in Appendix B.

\begin{table}[h]
\setlength{\tabcolsep}{9pt} 
\centering
\caption{Hallucinations for non-existent events evaluated in accuracy.  `Pos.' and `Neg.' refer to the model prediction accuracy of positive and negative queries, respectively. `/' denotes model hallucination, indicating the model is unable to determine whether the events in the query are present in the audio. Best results are in bold.}
\vspace{-0.15cm}
{
\begin{tabular}{lcc|cc}
\toprule
\multirow{2}{*}{Models} & \multicolumn{2}{c|}{Clotho-Moment }
& \multicolumn{2}{c}{AudioGrounding}  \\
\cline{2-5}
&   Pos. & Neg. &  Pos. & Neg. \\
\midrule
Kimi-Audio & 54.8 & 65.2 & 54.5 & 60.7  \\
TimeAudio  & / & / & / & /  \\
Qwen2-Audio & 72.2 & 43.1 & 57.6 & 55.1 \\
\rowcolor{blue!5} \textbf{\texttt{SpotSound-Q} }  & \textbf{87.6} & 85.8 & 93.2 & 80.7\\
Audio Flamingo 3  & 65.6 & 70.3 & 89.1 & 76.0  \\
\rowcolor{blue!5} \textbf{\texttt{SpotSound-A} }  & 85.4 & \textbf{88.1} & \textbf{93.4} & \textbf{87.9} \\
\bottomrule
\end{tabular}
}
\label{tab:hall}
\end{table}

\vspace{3pt} \noindent \textbf{Qualitative Results.}
As shown in Figure~\ref{fig:result}b, an example from the Clotho-Moment test set reveals distinct behavioral differences among the baselines. With the exception of TimeAudio, all models attempt to provide a binary existence judgment. However, Kimi-Audio and Qwen2-Audio yield incorrect predictions. TimeAudio fails to provide a relevant judgment entirely, producing hallucinatory, query-irrelevant responses. In contrast, both Audio Flamingo 3 and our \textbf{\texttt{SpotSound-A}} accurately determine the absence of the queried sound event, consistent with the ground truth.

\subsection{Two-stage Joint Assessment} 
\label{sec:twostage}
We compare the performance with existing models by combining two stages: determining the presence of a sound event and predicting the corresponding time window.

\vspace{3pt} \noindent \textbf{Metrics.}
We utilize the F1‑score to evaluate model performance. Specifically, $F_1 = \frac{2 \cdot TP}{2 \cdot TP + FP + FN}$, 
where TP, FP, and FN denote the number of true positives, false positives, and false negatives, respectively. The criteria for these metrics are defined through a two-stage evaluation process:
For negative queries (i.e., the target sound event is absent), predictions are categorized as true negatives (TN) if the model correctly identifies the absence of the event, and false positives (FP) otherwise.
For positive queries (i.e., the target sound event is present), a prediction is classified as a true positive (TP) if the first stage correctly detects the presence of the sound event and the second stage achieves an Intersection over Union (IoU) greater than 0.3. If the model fails to detect the event, it is considered a false negative (FN). Any predicted event that yields an IoU of 0.3 or lower is classified as a false positive (FP).

\vspace{3pt} \noindent \textbf{Results.}
As shown in Table~\ref{tab:two}, we can observe the following:
(i) TimeAudio fails to complete the two-stage evaluation due to the hallucinations.
(ii) Large ALMs underperform because of their susceptibility to hallucinating non-existent events and a lack of audio temporal grounding capabilities.
(iii) In contrast, our proposed model consistently maintains a highly competitive performance.

\begin{table}[t]
\setlength{\tabcolsep}{8.3pt} 
\centering
\caption{Two-stage joint assessment results evaluated in F1-score.  
`Clotho.', `UnAV.', `Spot.' and `Audio.' refer to Clotho-Moment, Unav-100 subset, \textbf{\texttt{SpotSound-Bench}} and AudioGrounding, respectively. Best results are in bold.}
\vspace{-0.15cm}
{
\begin{tabular}{lcccc}
\toprule
Models &   Clotho. & UnAV. &  Spot. & Audio.\\
\midrule
Kimi-Audio & 0.5 & 4.2 & 1.6 & 3.4 \\
TimeAudio  & / & / & / & /  \\
Qwen2-Audio & 5.4 & 11.1 & 3.4 & 41.6 \\
\rowcolor{blue!5} \textbf{\texttt{SpotSound-Q} } & \textbf{92.0} & \textbf{89.7} & 72.4 &  81.6 \\
Audio Flamingo 3  & 30.4 & 42.4 & 21.2 & 69.2 \\
\rowcolor{blue!5} \textbf{\texttt{SpotSound-A} }   & 91.4 & 84.9 & \textbf{82.9} & \textbf{85.6} \\
\bottomrule
\end{tabular}
}
\label{tab:two}
\end{table}

\subsection{Sound Event Detection}
\label{sec:sed}

In this section, we further assess the generalization of  \textbf{\texttt{SpotSound}} on event detection benchmarks.

\begin{table}[b]
\centering
\caption{Performance on sound event detection. Best results are in bold.}
\vspace{-0.15cm}
\setlength{\tabcolsep}{8.5pt} 
{
\begin{tabular}{lcc|cc}
\toprule
\multirow{2}{*}{Models} & \multicolumn{2}{c|}{TUT-Sound } & \multicolumn{2}{c}{DESED}  \\
\cline{2-5}
&  R1@.5 & mIoU &  R1@.5 & mIoU \\
\midrule
Kimi-Audio  & 0.3 & 1.5 & 4.0 & 4.1  \\
TimeAudio  & 18.0 & 22.5 & 4.0 & 19.3  \\
Qwen2-Audio  & 0.7 & 2.6 & 28.2 & 33.8 \\
\rowcolor{blue!5} \texttt{\textbf{SpotSound-Q}} & 23.0& 26.9 & \textbf{66.6} & \textbf{61.1}  \\
Audio Flamingo 3  & 6.7 & 17.2 & 51.8 & 53.7\\
\rowcolor{blue!5} \texttt{\textbf{SpotSound-A}} & \textbf{30.7}  & \textbf{33.2} & 58.0 &  57.8 \\
\bottomrule
\end{tabular}
}
\label{tab: sed}
\end{table}

\vspace{3pt} \noindent \textbf{Benchmarks.}
We evaluate on TUT Sound Events 2017~\cite{mesaros2016tut} and DESED~\cite{turpault2019sound,serizel2020sound}. For each dataset, we assemble test sets by aligning labels, timestamps, and audio recordings. Notably, as the original audio recordings in the TUT-Sound Events 2017 dataset are very long, we segment 
them into 60-second clips to facilitate processing.

\vspace{3pt} \noindent \textbf{Results.}
As revealed in Table~\ref{tab: sed}, we can make the following observations: (i) due to the length and complexity of TUT-Sound Events 2017, all models struggle, yet our models perform best with an mIoU of 26.9 and 33.2; (ii) audio clips in DESED are 10 seconds long, matching the training data distribution in most previous large ALMs and leading to higher temporal grounding accuracy, while our model still achieves the highest results.

\subsection{Ablation Study}
\label{sec:abla}

We conduct comprehensive ablation studies to evaluate the contribution of individual components and investigate the influence of various hyperparameters.

\vspace{3pt} \noindent \textbf{Ablation on Timestamp Interleaving. }
We ablate two key design choices across AudioGrounding, Clotho-Moment, UnAV-100, and \texttt{\textbf{SpotSound-Bench}}: (i) relaxing the 30-second encoder constraint to preserve long-context continuity, and (ii) interleaving timestamp tokens with audio tokens to provide explicit temporal grounding, applied to both \texttt{\textbf{SpotSound-Q}} and \texttt{\textbf{SpotSound-A}}. 

\begin{table}[h]
\setlength{\tabcolsep}{8.3pt} 
\centering
\caption{Ablation on different modules evaluated in mIoU.  
`Clotho.', `UnAV.', `Spot.' and `Audio.' refer to Clotho-Moment, Unav-100 subset, \textbf{\texttt{SpotSound-Bench}} and AudioGrounding, respectively. `(+) FT' denotes the standard fine-tuned version of baselines, `(+) unlock' denotes the Qwen2-Audio without the 30-second encoder limitation. Best results are in bold.}
\vspace{-0.15cm}
{
\begin{tabular}{lcccc}
\toprule
Models &   Clotho. & UnAV. &  Spot. & Audio.\\
\midrule
Qwen2-Audio & 5.7 & 9.7 & 4.4 & 37.0 \\
\quad (+) FT & 59.2 & 50.7 & 23.8 & 62.5  \\
\quad (+) unlock & 68.5 & 59.7 & 34.4 & 63.1  \\
\quad (+) timestamps  & \textbf{85.4} & \textbf{72.4 }& \textbf{ 49.9 }& \textbf{67.8}\\
\midrule
Audio Flamingo 3 & 22.6 & 25.0 & 7.6 & 47.5\\
\quad (+) FT & 82.8 & 52.7 & 37.4 & 60.3\\
\quad (+) timestamps  & \textbf{85.6} & \textbf{69.8 }& \textbf{ 57.9 }& \textbf{70.3}\\
\bottomrule
\end{tabular}
}
\label{tab:abla}
\end{table}

As shown in Table~\ref{tab:abla}, (i) comparing the fine-tuned version of Qwen2-Audio and fine-tuning Qwen2-Audio without the 30-second encoder limitation, we observe the effects of lifting the 30-second duration limit on the audio encoder. While this modification allows the model to ingest and perceive the full longer audio segments, thereby preserving complete semantic continuity, the resulting performance gain was relatively modest, yielding an improvement of  9.3\% on Clotho-Moment,  9.0\% on UnAV-100 subset,  10.6\% on \textbf{\texttt{SpotSound-Bench}}, 0.6\% on AudioGrounding; (ii) On both baselines, the introduction of interleaved absolute timestamps provides the critical temporal grounding missing in the previous configuration. This enhancement significantly sharpens the model's temporal resolution, leading to a substantial performance increase of  16.9/2.8\% on Clotho-Moment,  12.7/17.1\% on UnAV-100 subset,  15.5/20.5\% on \textbf{\texttt{SpotSound}},  4.7/10.0\% on AudioGrounding of two baselines.

\vspace{3pt} \noindent \textbf{Granularity of Timestamps. } 
The granularity at which timestamps are interleaved with audio tokens is a critical hyperparameter. As shown in Table~\ref{tab:param}, we evaluate three settings and find that for longer-form benchmarks (e.g., Clotho-Moment), coarsening the granularity to 2 seconds yields superior results, while short-clip benchmarks (e.g., AudioGrounding) benefit from a finer granularity of 0.2 seconds, albeit at the expense of increased training and inference time. Furthermore, latency evaluations in Appendix B.2
demonstrate that finer-grained timestamps incur higher inference latency. Consequently, to optimally balance between overall benchmark performance and computational efficiency, we set the timestamp granularity to 1 second.

\vspace{3pt} \noindent \textbf{Impact of AudioSet Strong Label and Synthetic Data.}
We investigate the impact of the mixing ratio between AudioSet Strong Label and synthetic samples, which governs the balance between real and synthetic data as well as between short- and long-form instances during training. AudioSet Strong Label provides reliable timestamps but is limited to 10-second clips; including too many such short samples degrades performance on long-form benchmarks. Conversely, weighting the mixture too heavily toward long-form synthetic data improves results on Clotho-Moment but hurts shorter benchmarks. To find the optimal balance, we evaluate three data mixing configurations comprising AudioSet Strong and synthetic samples, respectively: 5k and 10k, 10k and 10k, and 10k and 20k. As shown in Table~\ref{tab:param}, considering the trade-off between the performance across all four benchmarks and the computational cost associated with larger data volumes, we set the dataset to include 5k AudioSet and 10k synthetic samples.

\vspace{3pt} \noindent \textbf{Trainable Parameter Size. }
We conduct hyperparameter experiments on LoRA parameters, setting $r=8, 16, 32$ and $\alpha=2r$. The experimental results in Table~\ref{tab:param} show that the model achieves optimal performance when $r=8$ and $\alpha=16$.
We conduct these experiments based on \texttt{\textbf{SpotSound-A}}. Moreover, we provide additional ablation results of \texttt{\textbf{SpotSound-Q}} in the Appendix B.1.

\begin{table}[t]
\setlength{\tabcolsep}{9.5pt} 
\centering
\caption{Ablation results of hyperparameters evaluated in mIoU. The experiments are based on \texttt{\textbf{SpotSound-A}}. `Clotho.', `UnAV.', `Spot.' and `Audio.' refer to Clotho-Moment, Unav-100 subset, \textbf{\texttt{SpotSound-Bench}} and AudioGrounding, respectively. Settings in our experiments are in bold.}
\vspace{-0.15cm}
{
\begin{tabular}{lcccc}
\toprule
Settings &   Clotho. & UnAV. &  Spot. & Audio.\\
\midrule
\rowcolor{gray!10}   \multicolumn{5}{c}{\textit{Granularity of Timestamps}}\\
0.2s & 85.8 & 69.7 & 57.2 & 72.7 \\
1s & \textbf{85.6} & \textbf{69.8} & \textbf{57.9} & \textbf{70.3}  \\
2s & 87.2 & 69.6 & 56.5 & 69.7  \\

\midrule
\rowcolor{gray!10}   \multicolumn{5}{c}{\textit{Quantity of ASSL and Synthetic Data}}\\
5k\&10k & \textbf{85.6} & \textbf{69.8} & \textbf{57.9} & \textbf{70.3}  \\
10k\&10k & 84.7 & 72.3 & 57.4 & 71.7  \\
10k\&20k & 87.0 & 68.9 & 56.9 & 70.0  \\

\midrule
\rowcolor{gray!10}   \multicolumn{5}{c}{\textit{Trainable Parameters Size}}\\
$r=8, \alpha=16$ & \textbf{85.6} & \textbf{69.8} & \textbf{57.9} & \textbf{70.3} \\
$r=16, \alpha=32$& 85.5 & 67.8 & 57.3 & 69.5 \\
$r=32, \alpha=64$ & 82.6 & 60.6 & 47.3 & 64.0 \\
\bottomrule
\end{tabular}
\vspace{-0.25cm}
}
\label{tab:param}
\end{table}

%% file: sections/2_related_works.tex
\section{Related Work}
\label{sec:related}

\noindent \textbf{Large Audio Language Models.}
The development of Large Audio Language Models (ALMs) has shifted the field from task-specific systems towards unified audio-language understanding and generation assistants. Models such as the Qwen-Audio series~\cite{chu2023qwen, chu2024qwen2}, Audio Flamingo series~\cite{kong2024audio, ghosh2025audio, goel2025audio}, and Kimi Audio~\cite{ding2025kimi} exemplify this trend, leveraging large language models for versatile audio processing and reasoning. Despite these advances, a common limitation persists: these models exhibit weaker perception of environmental sounds compared to speech and music, and a pronounced deficiency in temporally localizing events within audio.

\vspace{3pt}\noindent \textbf{Audio Temporal Understanding.}
This research domain focuses on aligning language queries with specific audio segments. Initial efforts established the Text-to-Audio Grounding (TAG) task under full supervision~\cite{xu2021text}, followed by weakly-supervised paradigms (WSTAG) to reduce annotation costs~\cite{xu2024towards}. The scope expanded to long-form audio segment retrieval, prompting specialized architectures~\cite{munakata2025language}, while high-resolution datasets like AudioTime~\cite{xie2025audiotime} were introduced for precise temporal control. Recent work aims to integrate temporal reasoning as a core capability of ALMs for complex tasks like audio-grounded QA~\cite{sridhar2025enhancing}. TimeAudio~\cite{wang2025timeaudio} enables efficient long audio understanding via temporal markers and token merging. 
Existing methods focus on distinct, long-duration events, leaving the precise grounding of short, fleeting sounds within complex backgrounds underexplored; \textbf{\texttt{SpotSound}} addresses this.

\vspace{3pt}\noindent \textbf{Video Temporal Understanding.}
Video Temporal Grounding (VTG) tasks span both short and long videos, with short-video methods dominated by DETR-like architectures~\cite{carion2020end, lei2021detecting, gordeev2026saliency, moon2023correlation, moon2023query} and non-DETR approaches leveraging multi-modal cues~\cite{liu2022umt, boris2024surprising}. However, these struggle with long videos where relevant moments are sparse~\cite{hou2023cone, pan2023scanning}. The fundamental differences between short and long videos hinder unified models; while UniVTG~\cite{lin2023univtg} attempts to bridge this gap, its lightweight architecture limits generalization~\cite{shi2025enhancing}. Recent progress in Multi-modal Language Models (MLLMs) offers promise~\cite{lai2024lisa, pi2023detgpt, liu2025lamra}, but accurate temporal grounding remains challenging. Existing MLLM approaches fall into three paradigms: time-agnostic models lacking temporal signals~\cite{huang2024vtimellm, huang2024lita}, implicit timestamp-encoded models prone to hallucination~\cite{ren2024timechat, guo2025vtg}, and explicit temporal marking models constrained by context windows on long videos~\cite{boris2024surprising, chen2024timemarker, zhang2025videollama}. Unitime~\cite{li2025universal} introduces multi-scale coarse-to-fine grounding for long videos, while TimeLens~\cite{zhang2025timelens} establishes a robust data-driven baseline. 
Inspired by these advancements, 
\textbf{\texttt{SpotSound}} adapts explicit temporal marking to the audio domain, aiming to achieve precise, 
hallucination-free grounding in complex ``needle-in-a-haystack" scenarios.

%% file: sections/6_conclusion.tex
   \section{Conclusion}
\label{sec:conclusion}

In this paper, we introduced \textbf{\texttt{SpotSound}}, addressing the absence of precise temporal grounding in large audio–language models. Our approach combines a timestamp-interleaved alignment strategy with a training setup that explicitly mitigates hallucinations, enabling accurate localization of short acoustic events in continuous audio. To evaluate temporal acuity under realistic ``needle-in-a-haystack'' scenario, we released \textbf{\texttt{SpotSound-Bench}}, a curated benchmark emphasizing short-window events embedded in complex scenes. 
Across multiple benchmarks, \textbf{\texttt{SpotSound}} delivers state-of-the-art or highly competitive temporal grounding performance while maintaining strong results on sound event detection. These advances narrow the gap between coarse semantic understanding and fine-grained temporal reasoning, moving ALMs toward reliable use in real-world, time-critical applications.

%% file: sections/9_appendix.tex
\setcounter{figure}{0}
\setcounter{table}{0}
\renewcommand{\thefigure}{S\arabic{figure}}
\renewcommand{\thetable}{S\arabic{table}}

In the appendix, we provide dataset and benchmark statistics (Appendix~\ref{sec:statis}), more experiment results (Appendix~\ref{sec:more}), additional implementation details (Appendix~\ref{sec:add}), qualitative results (Appendix~\ref{sec:quality}), and discussions of limitations and future work (Appendix~\ref{sec:limit}).

\section{Dataset and Benchmark Statistics}
\label{sec:statis}
In this section, we conduct further statistics and analysis on datasets and benchmarks from multiple dimensions.

\subsection{Dataset Statistics}
\label{sec:statisd}
\begin{figure}[h]
\centering
\includegraphics[width=\linewidth]{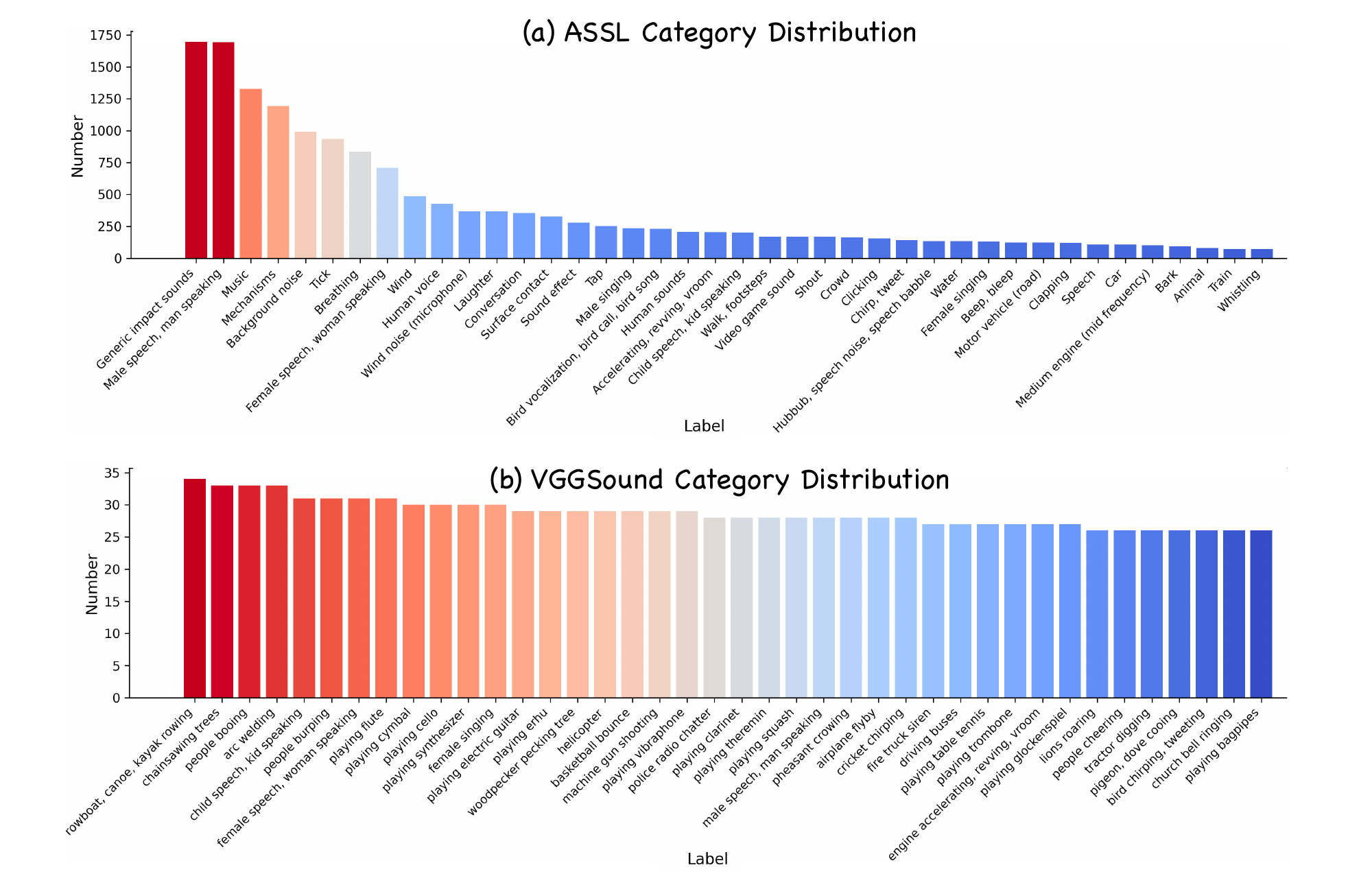}
\vspace{-0.6cm}
\caption{Category distributions for the synthetic dataset.}
\vspace{-0.3cm}
\label{fig:dataset}
\end{figure}

To construct our synthetic dataset, we curate a total of 10,000 audio-visual samples, randomly drawing 5,000 instances each from VGGSound~\cite{chen2020vggsound} and the AudioSet Strong Label (ASSL) dataset~\cite{hershey2021benefit}. As depicted in Figure~\ref{fig:dataset}, our statistical analysis confirms that this sampling strategy faithfully preserves the underlying class priors of the source distributions. Notably, the ASSL subset exhibits a natural skew toward high-frequency anthropogenic classes—such as human speech and generic impact sounds—while the VGGSound subset contributes a comparatively uniform semantic distribution.

\begin{figure}[h]
\centering
\includegraphics[width=\linewidth]{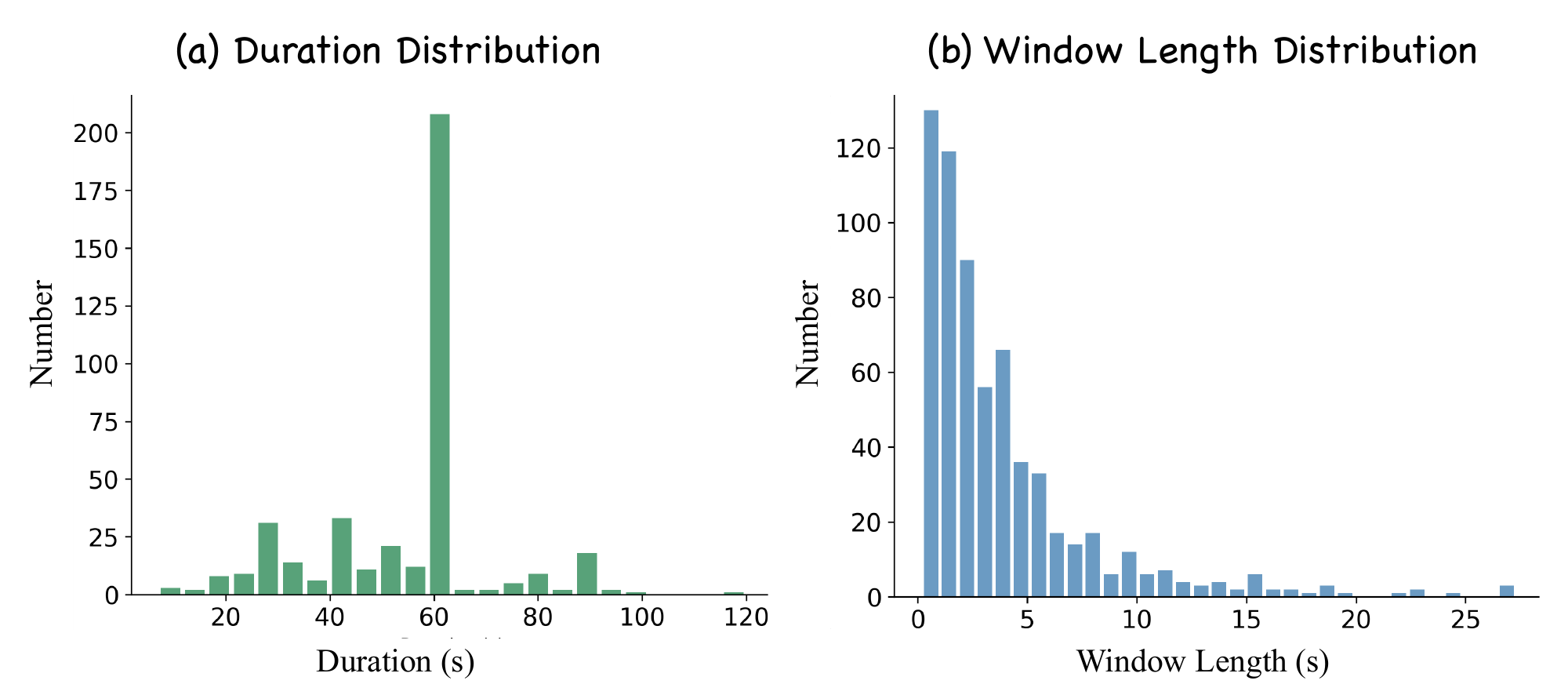}
\vspace{-0.6cm}
\caption{Duration and window length distributions for \texttt{\textbf{SpotSound-Bench}}.}
\vspace{-0.3cm}
\label{fig:bench}
\end{figure}
\subsection{Benchmark Statistics}
\label{sec:statisb}

To characterize the temporal dynamics of \texttt{\textbf{Spotsound-Bench}}, we analyze the distributions of both the full sample durations and the annotated event window lengths. As illustrated in Figure~\ref{fig:bench}, the benchmark predominantly features untrimmed videos with durations centered around 60 seconds, whereas the target audio-visual events are highly localized, with the vast majority of window lengths constrained to the 0 to 10-second range. This pronounced disparity between the global video length and the local event duration is a critical feature of our dataset. It effectively simulates a realistic, sparse temporal localization problem, a "needle-in-a-haystack" scenario—thereby rigorously challenging models to demonstrate both robust long-range temporal context modeling and fine-grained temporal grounding capabilities.

\section{More Experiment Results}
\label{sec:more}
In Section~\ref{sec:abla}, 
we conduct a series of ablation experiments based on \texttt{\textbf{SpotSound-Q}}. In Section~\ref{sec:latency}, we measured the runtime consumption of ALMs. In Section~\ref{sec:robust}, we performed robustness testing on the model. In Section~\ref{sec:halluss}, we present supplementary experimental results on hallucination for negative samples on the \texttt{\textbf{SpotSound-Bench}} and UnAV-100 subset.

\subsection{Ablation Study of \texttt{\textbf{SpotSound-Q}}}
\label{sec:paramq}
Building upon the architectural investigations of \texttt{\textbf{SpotSound-A}} in Section~\ref{sec:abla}, we conduct a parallel suite of ablation studies for \texttt{\textbf{SpotSound-Q}}. Specifically, we evaluate the mean Intersection over Union (mIoU) to assess the impact of three core factors: (i) timestamp granularity, (ii) the integration of AudioSet Strong Label (ASSL) and synthetic training data, and (iii) the capacity of trainable parameters.

\begin{table}[h]
\setlength{\tabcolsep}{9.5pt} 
\centering
\caption{Ablation Results of Hyperparameters.  `Clotho.', `UnAV.', `Spot.' and `Audio.' refer to Clotho-Moment, Unav-100 subset. Settings in our experiments are in bold.}
{
\begin{tabular}{lcccc}
\toprule
Settings &   Clotho. & UnAV. &  Spot. & Audio.\\
\midrule
\rowcolor{gray!10}   \multicolumn{5}{c}{Granularity of Timestamps}\\
\midrule
0.2s & 82.6 & 69.9 & 42.3 & 61.2 \\
1s & \textbf{85.4} & \textbf{72.4} & \textbf{46.6} & \textbf{67.8} \\
2s & 81.5 & 70.1 & 38.7 & 60.5 \\

\midrule
\rowcolor{gray!10}   \multicolumn{5}{c}{Quantity of ASSL and Synthetic Data}\\
\midrule
5k:10k &  \textbf{85.4} & \textbf{72.4} & \textbf{46.6} & \textbf{67.8} \\
10k:10k & 84.9 & 70.2 & 44.1 & 66.7 \\
10k:20k & 86.7 & 67.6 & 45.8 & 67.1 \\

\midrule
\rowcolor{gray!10}   \multicolumn{5}{c}{Trainable Parameters Size}\\
\midrule
$r=8, \alpha=16$ &  \textbf{85.4} & \textbf{72.4} & \textbf{46.6} & \textbf{67.8} \\
$r=16, \alpha=32$& 86.8 & 69.4 & 42.7 & 62.3 \\
$r=32, \alpha=64$ & 82.7 & 63.9 & 39.7 & 55.9 \\
\bottomrule
\end{tabular}
}
\label{tab:paramq}
\end{table}

\vspace{3pt} \noindent \textbf{Results.} As detailed in Table~\ref{tab:paramq}, \texttt{\textbf{SpotSound-Q}} yields peak performance across all four benchmarks under the following configuration: (i) a timestamp granularity of 1 second, which optimally balances fine-grained temporal resolution with sequence length constraints; (ii) a hybrid training corpus comprising 5k ASSL and 10k synthetic samples; and (iii) a parameter-efficient LoRA fine-tuning setup utilizing rank $r=8$ and scaling factor $\alpha=16$, which ensures sufficient representational expressivity while mitigating the risk of overfitting.

\subsection{Latency in Inference Process}
To assess the computational efficiency of our framework, we evaluate the inference latency under two distinct timestamp granularity configurations, as previously introduced in Section~\ref{sec:abla}. Because the chosen temporal resolution directly dictates the length of the generated token sequence, it serves as a primary determinant of inference speed. Specifically, we benchmark both the initial model loading overhead and the total inference time required to process a standardized subset of 100 samples from \texttt{\textbf{SpotSound-Bench}}.
\label{sec:latency}
\begin{table}[h]
\setlength{\tabcolsep}{7.5pt} 
\centering
\caption{Latency in Inference Process. Settings in our experiments are in bold.}
{
\begin{tabular}{lcc}
\toprule
&  Load Model & Inference (1 Sample) \\
\midrule
 \textbf{\texttt{SpotSound-A} }  & \textbf{7.6s} & \textbf{1.0s} \\
\textbf{\texttt{SpotSound-A} } - 0.2s & 7.5s & 1.4s\\
\textbf{\texttt{SpotSound-A} } - 2s & 7.6s & 1.0s \\
\bottomrule
\end{tabular}
}
\label{tab:latency}
\end{table} 
\vspace{3pt} \noindent \textbf{Results.} As detailed in Table~\ref{tab:latency}, increasing the temporal resolution (i.e., employing finer timestamp granularity) incurs a noticeable penalty in inference latency. From an architectural standpoint, this trade-off is expected: finer granularity necessitates the autoregressive generation of a significantly larger number of textual timestamp tokens. This is attributed to the fact that finer granularity necessitates more frequent insertion of textual timestamps, thereby incurring greater time overhead.

\subsection{Robustness Test}
\label{sec:robust}
Generative models are prone to producing hallucinations in their outputs. To assess robustness, we conduct experiments on the SpotSound benchmark from two perspectives: (i) perturbing the temporal positions of events to evaluate robustness against temporal distribution bias, and (ii) paraphrasing fixed category queries into synonyms to test model reliability under varying query formulations. 
\begin{table}[h]
\setlength{\tabcolsep}{9.5pt} 
\centering
\caption{Robustness test results. `Target' and denotes randomly shifting the target sound event window. Settings in our experiments are in bold.}
\vspace{-0.15cm}
{
\begin{tabular}{lcccc}
\toprule
Settings &   R1@.3 & R1@.5 &  R1@.7 & mIoU\\
\midrule
\textbf{\texttt{SpotSound-A}} & \textbf{76.5}&  \textbf{59.9}&  \textbf{40.0} & \textbf{57.9} \\
\midrule
\rowcolor{gray!10}   \multicolumn{5}{c}{\textit{Temporal Event Shifting}}\\
Target  & 72.3 & 51.0 & 37.8 & 53.2\\

\midrule
\rowcolor{gray!10}   \multicolumn{5}{c}{\textit{Query Paraphrasing}}\\
Synonyms &75.0 & 57.8 & 39.5 & 56.3\\
Questions & 74.8 & 55.0 & 37.8 & 54.9 \\
\bottomrule
\end{tabular}
}
\vspace{-0.1cm}
\label{tab:robust}
\end{table}

\vspace{3pt} \noindent \textbf{Temporal Event Shifting.}
Prior literature highlights that temporal localization datasets frequently suffer from severe event-distribution biases, allowing models to exploit statistical positional priors rather than performing genuine grounding~\cite{hao2022can, otani2020uncovering, li2025universal}. To rigorously evaluate whether our model relies on such spurious correlations, we introduce a temporal perturbation strategy, detailed in Table~\ref{tab:robust}. Specifically, we extract the target sound event and randomly re-insert it into a background context, thereby neutralizing any dataset-specific temporal priors. Under this strict evaluation setting, our method exhibits remarkable resilience, maintaining high localization accuracy. This confirms that our model's performance stems from robust acoustic-semantic understanding rather than the exploitation of superficial temporal shortcuts.

\vspace{3pt} \noindent \textbf{Query Paraphrasing.} 
Genuine temporal grounding demands a deep semantic alignment between the audio stream and the text prompt, rather than brittle lexical matching against fixed query templates. To evaluate our model's semantic robustness and resistance to prompt fragility, we employ DeepSeek-v3~\cite{liu2024deepseek} to paraphrase the standard \textbf{\texttt{SpotSound-Bench}} queries into distinct linguistic variations, as detailed in Table~\ref{tab:robust}. Specifically, we generate synonymous rewrites (Synonyms) and interrogative reformulations (Questions). For each paraphrased query, we measure the temporal Intersection over Union (IoU) between the predicted boundaries and the ground-truth segments. As demonstrated in Table~\ref{tab:robust}, our method exhibits high resilience to these linguistic perturbations, maintaining accurate and reliable localization across both synonymous and interrogative rephrasings. This confirms that our model successfully captures the underlying acoustic-semantic concepts rather than merely memorizing specific text prompts.

\subsection{Hallucination for Negative Samples}
\label{sec:halluss}
Expanding upon the hallucination analysis presented in Section~\ref{sec:hallu}, we further investigate the susceptibility of ALMs to hallucinate non-existent events. Here, we provide supplementary evaluations on \texttt{\textbf{SpotSound-Bench}} and the UnAV-100 subset~\cite{geng2023dense}.

\begin{table}[h]
\setlength{\tabcolsep}{8pt} 
\centering
\caption{Hallucinations for non-existent events evaluated in accuracy.  `Pos.' and `Neg.' refer to the model prediction accuracy of positive and negative queries, respectively. `/' denotes model hallucination, indicating the model is unable to determine whether the events in the query are present in the audio. Best results are in bold.}
\vspace{-0.15cm}
{
\begin{tabular}{lcc|cc}
\toprule
\multirow{2}{*}{Models} & \multicolumn{2}{c|}{\texttt{\textbf{SpotSound-Bench}}}
& \multicolumn{2}{c}{UnAV-100 subset}  \\
\cline{2-5}
&   Pos. & Neg. &  Pos. & Neg. \\
\midrule
Kimi-Audio & 51.8 & 60.5 & 50.0 & 41.0  \\
TimeAudio  & / & / & / & /  \\
Qwen2-Audio & 55.5 & 78.0 & 64.0 & 44.0  \\
\rowcolor{blue!5} \textbf{\texttt{SpotSound-Q}}  & 85.3 & 88.3 & 93.0 & \textbf{96.0 } \\
Audio Flamingo 3  & 85.5 & 95.0 & 80.0 & 92.0  \\
\rowcolor{blue!5} \textbf{\texttt{SpotSound-A}} & \textbf{92.3} & \textbf{96.5} & \textbf{94.0 }& 93.0  \\
\bottomrule
\end{tabular}
}
\vspace{-0.2cm}
\label{tab:hall_ap}
\end{table}
\vspace{3pt} \noindent \textbf{Results.} As detailed in Table~\ref{tab:hall_ap}, TimeAudio suffers from severe prior-induced hallucination, frequently failing to distinguish between the actual presence and the absence of queried sound events. While other contemporary ALMs exhibit marginal robustness against such false positives, their discriminative performance remains suboptimal. Conversely, our proposed model establishes a new state-of-the-art across all four evaluation metrics on both benchmarks. 

\section{Additional Implementation Details}
\label{sec:add}
We provide the prompt template for querying \textbf{\texttt{SpotSound}} and the prompt used for generating audio captions of foreground sounds in the synthesis data process.

\subsection{Prompt Template for \texttt{\textbf{SpotSound}}}
As formulated in Section~\ref{sec:problem}, we decompose the temporal grounding task into a decoupled, two-stage inference pipeline. Rather than directly predicting timestamps—which often exacerbates hallucination—our model must first explicitly verify the existence of the queried sound event within the audio-visual stream. Only upon a positive detection does the model proceed to the second stage: precise temporal localization. To facilitate this structured reasoning process, we design the following task-specific prompt templates for each stage:
\begin{casestudy}{Prompt for Sound Event Identification}

System: You are a helpful assistant.

\medskip

User: This is a sequence of audio stream. Your task is to identify whether the sound event in the query occurs. The query is: \{query\}.

\end{casestudy}

\begin{casestudy}{Prompt for Audio Temporal Grounding}

System: You are a helpful assistant.

\medskip

User: This is a sequence of audio stream. Your task is to identify the temporal window (start and end timestamps) when the given query appears. The query is: \{query\}.

\end{casestudy}

\subsection{Prompt for Foreground Query Generation}
As detailed in Section~\ref{sec:data}, we construct a synthetic dataset comprising 10,000 samples by leveraging AudioSet Strong Label (ASSL) and VGGSound as foreground audio events. To ensure high-quality and contextually rich textual annotations for this synthesized data, we employ an automated captioning pipeline driven by state-of-the-art foundation models: Qwen2-Audio~\cite{chu2024qwen2} and DeepSeek-v3~\cite{liu2024deepseek}.

Specifically, to extract detailed acoustic descriptions directly from the raw VGGSound clips, we process the audio streams through Qwen2-Audio. The prompt designed to elicit these fine-grained audio captions is structured as follows:

\begin{casestudy}{Prompt for Qwen2-Audio}

You are a professional audio captioner. 

\medskip

I will provide an audio stream. Please write a 10–15 word audio caption that fully and accurately describes the events in the audio, without any extraneous information. The audio stream is: \{audio\}.

\end{casestudy}

For the ASSL dataset, the raw annotations consist of discrete class labels and precise temporal boundaries. To convert this structured metadata into the natural language format required for effective ALM training, we prompt DeepSeek-v3 to synthesize these discrete events into a cohesive, chronologically accurate audio narrative. The prompt utilized for this text-to-text transformation is structured as follows:
\begin{casestudy}{Prompt for DeepSeek-v3}

You are a professional audio captioner. 

\medskip

I will provide you with several sound event labels and their start and end times within an audio stream. Based on these labels and timestamps, please write a 10–15 word audio caption that fully and accurately describes what happens in the audio, without any extraneous information. The labels and timestamps are: \{labels and timestamps\}.

\end{casestudy}

\section{Qualitative Results}
\label{sec:quality}
We present success and failure cases from the results of \texttt{\textbf{SpotSound-A}} on different benchmarks, and further analyse the ability of our model.

\subsection{Qualitative Results for \texttt{\textbf{SpotSound-Bench}}}

\begin{figure}[t]
\centering
\includegraphics[width=\linewidth]{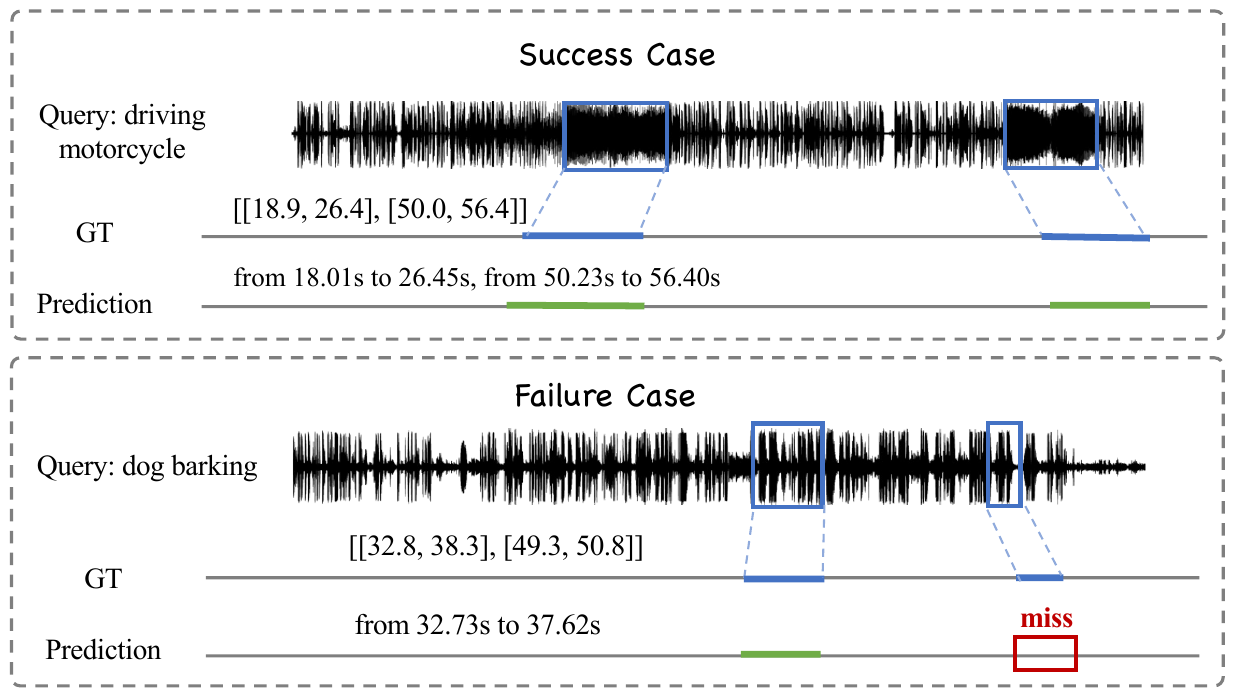}
\vspace{-0.6cm}
\caption{Success and failure cases from \texttt{\textbf{SpotSound-Bench}}.}
\vspace{-0.3cm}
\label{fig:spotsound}
\end{figure}

\texttt{\textbf{SpotSound-Bench}} presents a rigorous localization challenge, characterized by extended audio sequences that contain highly transient, short-duration sound events. As illustrated in Figure~\ref{fig:spotsound}, our model demonstrates robust temporal grounding capabilities, successfully isolating these sparse events in the majority of cases. However, we observe a specific failure mode in multi-instance scenarios: when a target sound event occurs across multiple distinct time windows within the same audio clip, the model occasionally fails to detect the complete set of occurrences. This limitation likely stems from the autoregressive decoding process, where the model may prematurely terminate generation after identifying the most salient instance, thereby overlooking secondary temporal windows.

\subsection{Qualitative Results for AudioGrounding}
\begin{figure}[h]
\centering
\includegraphics[width=\linewidth]{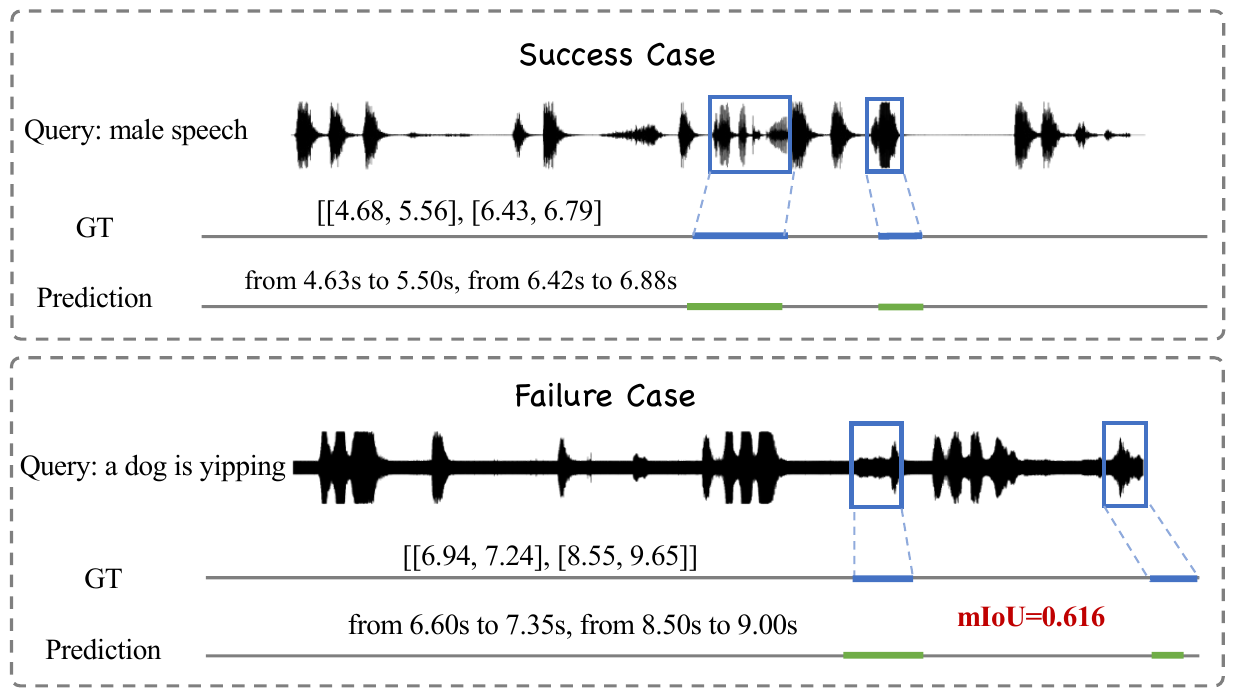}
\vspace{-0.6cm}
\caption{Success and failure cases from AudioGrounding.}
\vspace{-0.3cm}
\label{fig:ag}
\end{figure}

The AudioGrounding benchmark comprises short, 10-second audio clips. As illustrated in Figure~\ref{fig:ag}, while our model demonstrates robust overall localization, it occasionally exhibits slight boundary misalignments at a fine-grained level (e.g., predicting 9.65s instead of the ground-truth 9.00s). Although these absolute temporal errors are marginal in human perception, they disproportionately penalize the mean Intersection over Union (mIoU) metric, as the overlap ratio is highly sensitive to boundary precision when the target event's duration is extremely brief.

\subsection{Qualitative Results for UnAV-100 subset}
\begin{figure}[h]
\centering
\includegraphics[width=\linewidth]{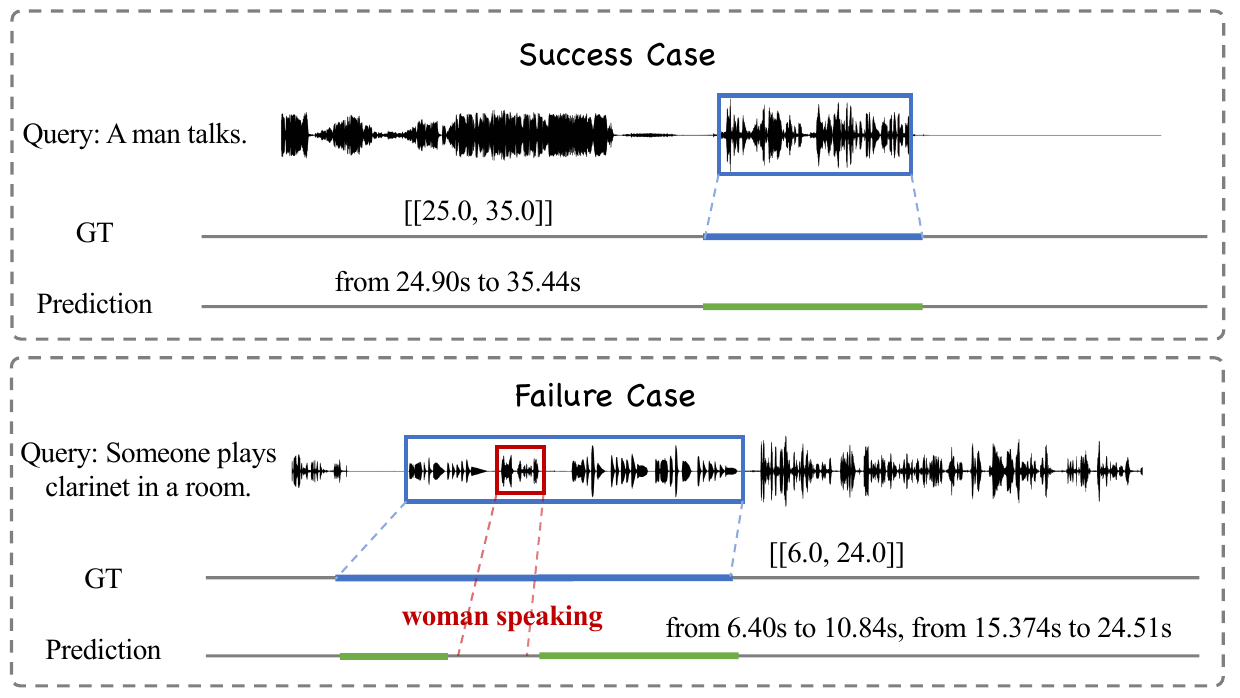}
\vspace{-0.6cm}
\caption{Success and failure cases from UnAV-100 subset.}
\vspace{-0.3cm}
\label{fig:unav}
\end{figure}

The UnAV-100 subset is characterized by sound events with extended durations. Notably, as shown in Figure~\ref{fig:unav} we observe that the ground-truth annotations in this benchmark often exhibit coarse temporal granularity; for instance, consecutive occurrences of the same sound class are frequently merged into a single, continuous temporal window, absorbing the silent intervals between them. Because our model possesses high fine-grained temporal resolution, it accurately detects these silent gaps and correctly segments the instances into distinct temporal windows. Consequently, while our model's predictions are acoustically more precise, this superior resolution leads to a structural deviation from the benchmark's coarse ground truth, resulting in an artificial penalty in the automated evaluation metrics.

\section{Limitation and Future Work}
\label{sec:limit}
Despite these promising results, our current framework exhibits a few notable limitations that pave the way for future research. First, while \textbf{\texttt{SpotSound}} excels at localizing distinct, sustained acoustic events, its temporal precision and generalization capabilities on short-window benchmarks (i.e., highly transient sounds) remain a bottleneck. Second, the model's localization accuracy is inherently bounded by the granularity and quality of the temporal annotations in the training corpus. Consequently, achieving greater robustness will require scaling the training pipeline with larger datasets that feature dense, fine-grained, and challenging acoustic samples.

Moving forward, our research will focus on audio temporal grounding in highly complex, real-world acoustic scenes. Specifically, we aim to tackle the challenges of polyphonic environments, where multiple distinct sound events overlap simultaneously, as well as improving multi-instance localization to ensure the accurate detection of all temporal windows when an event occurs repeatedly. In summary, while our current framework establishes a strong baseline for audio grounding, addressing these constraints in temporal resolution and complex scene understanding is the crucial next step toward fully robust real-world deployment.